\documentclass[%
preprint,
amsmath,amssymb,
prb,
]{revtex4-2}

\usepackage{graphicx}% Include figure files
\usepackage{dcolumn}% Align table columns on decimal point
\usepackage{bm}% bold math
\usepackage{hyperref}% add hypertext capabilities
\usepackage{siunitx}
\usepackage{lineno}
\DeclareSIUnit\rydberg{Ry}

\begin{document}
	
	% \preprint{APS/123-QED}
	
%	\linenumbers
	\title{Superconductivity in Nb: from the impact of temperature, Cooper-pairing to dimensionality}% Force line breaks with \\
	
	\author{Uriel A. Aceves Rodriguez$^{1,2}$}
	\author{Filipe Guimar\~aes${^3}$}%
	\author{Samir Lounis$^{1,2}$}%
	\email{s.lounis@fz-juelich.de}
	\affiliation{$^{1}$Peter Gr\"unberg Institut and Institute for Advanced Simulation, Forschungszentrum J\"ulich \& JARA, 52425 J\"ulich, Germany\\
		$^{2}$Faculty of Physics \& CENIDE, University of Duisburg-Essen, 47053 Duisburg, Germany\\
		$^{3}$J\"{u}lich Supercomputing Centre, Forschungszentrum J\"{u}lich \& JARA, 52425 J\"{u}lich, Germany
	}
	
	\date{\today}% It is always \today, today,
	
	
	\begin{abstract}
		The ability to simulate realistically the electronic structure of superconducting materials is important to understand and predict various properties emerging in both the superconducting topological and spintronics realms. We introduce a tight-binding implementation of the Bogoliubov-de Gennes method, parameterized from density functional theory, which we utilize to explore the bulk and thin films of Nb, known to host a significant superconducting gap. The latter is useful for various applications such as the exploration of trivial and topological in-gap states. Here, we focus on the simulations aspects of superconductivity and study the impact of temperature, Cooper-pair coupling and dimensionality on the size of the superconducting order parameter.
	\end{abstract}
	
	\maketitle

	\section{Introduction}
	
	Superconductivity is a basic fundamental phenomenon in solid-state physics. Although being discovered more than a century ago, it still generates a plethora of research activities. Recently, its importance flourished again in the context of topological superconductivity\cite{SciPostPhys.3.3.021,PRXQuantum.2.040347,PhysRevB.105.075129}, which is an essential field of research for the realization of quantum computing. Certainly, the intricate interplay of standard spintronics and quantum information concepts can give rise to unforeseen breakthroughs in basic research and in future technologies for storing, computing and transmitting information. This calls for the development of theoretical frameworks based on a realistic description of the electronic structure of superconducting materials interfaced with magnetic systems in order to understand and predict various related phenomena.
	
	Interesting topics emerged in superconducting spintronics~\cite{Linder2015,Eschrig2015,Buzdin2005,Bergeret2018}, where the interplay of Cooper pairs and other electronic degrees of freedom (such as spin, charge and orbital) give rise to fascinating phenomena, ranging from supercurrents, triplet superconductivity, and torques. Simultaneously, the potential design of topological superconductivity and the creation of Majorana states\cite{doi:10.1146,Kitaev_2001,Tanaka2009} is driving a lot of research in the context of topological qubits~\cite{Aguado2020}. 
	
	The microscopic theory to describe conventional superconductivity goes back to the work of John Bardeen, Leon Cooper, and Robert Schrieffer (BCS)\cite{PhysRev.106.162}. In this context, the Bogoliubov--de Gennes (BdG) method\cite{bogol,degennes,Alaberdin1996,zhu2016bogoliubov} is an elegant mean-field approximation that relies upon Bogoliubov--Valatin transformations that take the Hamiltonian from a particle space into a particle-hole one---a framework that has been frequently used, see for example Refs.~\citenum{Han_2009,PhysRevB.102.245106,Pellegrino2022,PhysRevB.105.L140504,Sato2017,Feldman2016,Schneider2020,Schneider2021,Schneider2022,doi:10.1126/sciadv.abi7291,Kster2022}. 
	Nowadays, several methods based on a realistic description of the electronic structure of materials are capable of treating fundamental aspects related to superconductivity. For instance, powerful theoretical frameworks have been proposed and utilized within density functional theory (DFT) combined with the BdG formalism\cite{PhysRevB.105.125143,Beck2021,PhysRevB.104.245415,Laszloffy2023,Bendeguz2023,Ruessmann2023}, or by proposing an extension of DFT to account for superconductivity~\cite{Oliveira1988,Lueders2005,Marques2005,Essenberger2016}. Tight-binding schemes are very appealing since they allow a versatile description of superconductivity-related physics with a potential control of various levels of approximations, which could enable the treatment of problems not attainable with conventional DFT.   
	
	Among the superconducting materials, Nb stands out and is often utilized as a key component to investigate superconductivity-related physics~\cite{Khaydukov2018,Pal2022}. The surface of Nb(110) became recently the standard substrate~\cite{PhysRevB.99.115437}, where Yu-Shiba-Rusinov in-gap states\cite{Kitaev_2001,doi:10.1146,Tanaka2009} and the topological Majorana states\cite{doi:10.1146,Kitaev_2001,Tanaka2009} are explored in adatoms\cite{Kster2021,felix,Beck2021,PhysRevB.102.174504}, at the edges of magnetic wires\cite{PhysRevB.103.235437,Schneider2021,doi:10.1126/sciadv.abi7291,Kster2022}, or films~\cite{Soldini2023,doi:10.1021/acsnano.2c03965,PhysRevMaterials.6.024801,PhysRevB.105.L100406}.
	
	The goal of this article is to introduce our recently developed method that allows the treatment of superconductivity on the basis of a tight-binding approach as implemented in TITAN~\cite{titan_zenodo,Guimaraes2017,Guimaraes2019,Guimaraes2020}. The framework enables a realistic description of the electronic structure, with parameters obtained from DFT~\cite{papabook}. Examples of applications of TITAN ranges from dynamical magnetic responses\cite{Guimaraes2015}, dynamical transport and torques\cite{Guimaraes2017,Guimaraes2020}, magnetic damping~\cite{Guimaraes2019}, as well as ultrafast magnetization dynamics triggered by laser pulses~\cite{Hamamera2022,Hamamera2023}. In the context of superconductivity, the method was already introduced in Ref.~\citenum{Aceves2023}, with a focus on magnetic exchange interactions at superconducting-magnetic interfaces. Here, we pay attention to Nb in bulk and thin films with the aim of highlighting fundamental aspects in the simulations procedure. Aspects related to the choice of the electron-phonon coupling to open the appropriate band gap are discussed together with the impact of temperature and self-consistency on the underlying physics. Dimensionality affects superconductivity, as we unveil, which can be utilized to tune the superconducting gap for different applications in modern investigations of superconductivity-related problems. 
	
	The article is organized as follows. We first introduce the method in Sec.~\ref{sec:bdg-theory}, which is based on multi-orbital tight-binding theory. Then we present in Sec.~\ref{sec:results} the results of our simulations in Nb bulk, (001) and (110) monolayers, which are followed by Nb(110) thin films of different thicknesses. Finally, we present our concluding remarks in Sec.~\ref{sec:conclusions}.

	\section{Tight-binding and the Bogoliubov--de Gennes method}\label{sec:bdg-theory}
	To simulate the superconducting properties of Nb across dimensions, we introduce here the tight-binding methodology together with the BdG method that we utilized, as implemented in our code TITAN~\cite{titan_zenodo,Guimaraes2019,Guimaraes2020,Hamamera2022,Hamamera2023,Aceves2023}. The hopping parameters were obtained from first-principles calculations~\cite{papabook}.  More details on superconducting related aspects are elaborated in Ref.~\citenum{Aceves2023}.
	
	The fundamental tight-binding Hamiltonian reads: 
	\begin{equation}
		\begin{aligned}
			H_{S} = \frac{1}{N}\left\{\sum_{\alpha\beta,\sigma \eta, \mu \nu,\bm{k}} H_{\alpha\beta,\sigma \eta}^{\mu \nu}(\bm{k}) c^{\dagger}_{\alpha\mu\sigma}(\bm{k}) c_{\beta\nu\eta}(\bm{k}) - \sum_{\alpha,\mu\nu,\bm{k}\bm{k}'} \lambda_{\alpha\mu} c^{\dagger}_{\alpha\mu\uparrow}(\bm{k}) c^{\dagger}_{\alpha\mu\downarrow}(-\bm{k}) c_{\alpha\mu\downarrow}(-\bm{k}') c_{\alpha\mu\uparrow}(\bm{k}')\right\}\,,
		\end{aligned}
		\label{eq:bcs}
	\end{equation}
	where $c^{\dagger}_{\alpha\mu\sigma}(\bm{k})$ and $c_{\beta\nu\eta}(\bm{k})$ consist of the creation and annihilation operators associated to electrons having a wave vector $\bm{k}$ and spin $\sigma$ in the orbitals $\mu$  and spin $\eta$ in the orbitals $\nu$ , respectively. $\alpha$  and $\beta$ are specific atoms in the bulk unit cell, or layers in the film geometry. $\bm{k}$ is a reciprocal vector, while  $N$ is the number of wave vectors in the three-dimensional (for bulk) or two-dimensional (for films) Brillouin zone. To enable the formation of Cooper pairs, which ultimately gives rise to superconductivity, we introduce  $\lambda_{\alpha\mu}$ in the second term of the previous equation, as induced by the electron-phonon coupling. 
	
	In the mean-field approximation, Eq.~\eqref{eq:bcs} simplifies to
	\begin{equation}
		\begin{aligned}
			H_{S}^{\text{MF}} = &\frac{1}{N}\sum_{\bm{k}}\left\{\sum_{\alpha\beta,\sigma \eta, \mu \nu} H_{\alpha\beta,\sigma\eta}^{\mu \nu}(\bm{k}) c^{\dagger}_{\alpha\mu\sigma}(\bm{k}) c_{\beta\nu\eta}(\bm{k})\right.\\ &- \left.\sum_{\alpha,\mu} \left(\Delta_{\alpha\mu}^{*} c_{\alpha\mu\downarrow}(-\bm{k})c_{\alpha\mu\uparrow}(\bm{k}) +  \Delta_{\alpha\mu} c_{\alpha\mu\uparrow}^\dagger(\bm{k}) c_{\alpha\mu\downarrow}^\dagger (-\bm{k}) \right)\right\},
		\end{aligned}
		\label{eq:mf-bcs}
	\end{equation}
	with
	\begin{equation}
		\Delta_{\alpha\mu} = \lambda_{\alpha\mu} \frac{1}{N}\sum_{\bm{k}}\langle c_{\alpha\mu\downarrow}(-\bm{k}) c_{\alpha\mu\uparrow}(\bm{k}) \rangle, \quad \Delta_{\alpha\mu}^{*} = \lambda_{\alpha\mu} \frac{1}{N}\sum_{\bm{k}}\langle c_{\alpha\mu\uparrow}^\dagger(\bm{k}) c_{\alpha\mu\downarrow}^\dagger(-\bm{k}) \rangle,
		\label{eq:delta_lambda}
	\end{equation}
	$\Delta_{\alpha\mu}$ is known as the superconducting gap parameter, which corresponds to the energy required to scatter Cooper pairs~\cite{tinkham2004introduction}. 
	
	The term $H_{\alpha\beta,\sigma \eta}^{\mu \nu}(\bm{k})$ corresponds to the regular Hamiltonian where superconductivity is not accounted for:
	\begin{equation}
		H^{\mu \nu}_{\alpha\beta,\sigma \eta}(\bm{k}) = H_{\alpha\beta}^{0\mu \nu}(\bm{k}) \sigma^0 + \bm{\sigma}\cdot \hat{\bm{e}}_\alpha B_\alpha^{[\text{xc}]\mu\nu}(\bm{k}) \delta_{\alpha\beta} + \bm{\sigma}\cdot \bm{B}_\alpha^{[\text{soc}]\mu\nu}(\bm{k})\delta_{\alpha\beta}\,,
	\end{equation}
	with $H_{\alpha\beta}^{0\mu \nu}$ being the spin-independent tight-binding term, the second term is responsible for the intra-atomic exchange interaction  (originating from a Hubbard-like contribution\cite{Guimaraes2017,Guimaraes2020}), which can lead to magnetism, and the third term corresponds to the spin-orbit interaction.

	The BdG methodology consists in utilizing the Bogoliubov--Valatin transformation\cite{zhu2016bogoliubov} to rewrite Eq.~\eqref{eq:mf-bcs} in the electron-hole representation instead of the electron one, which leads to the BdG Hamiltonian:
	\begin{equation}
		H_{\text{BdG}}^{\alpha\beta,\mu\nu}(\bm{k}) =\begin{pmatrix}
			H_{\alpha\beta,\uparrow\uparrow}^{\mu\nu}(\bm{k})-E_F & H_{\alpha\beta,\uparrow\downarrow}^{\mu\nu}(\bm{k})  & 0 & -\Delta_{\alpha\mu}
			\mathbb{I}\\ 
			H_{\alpha\beta,\downarrow\uparrow}^{\mu\nu}(\bm{k}) & H_{\alpha\beta,\downarrow\downarrow}^{\mu\nu}(\bm{k})-E_F  & \Delta_{\alpha\mu}
			\mathbb{I}& 0\\ 
			0 & \Delta_{\alpha\mu}^*
			\mathbb{I}& -H_{\alpha\beta,\uparrow\uparrow}^{\mu\nu *}(-\bm{k}) + E_F &-H_{\alpha\beta,\uparrow\downarrow}^{\mu\nu *}(-\bm{k})  \\ 
			-\Delta_{\alpha\mu}^*
			\mathbb{I}& 0 & -H_{\alpha\beta,\downarrow\uparrow}^{\mu\nu *} (-\bm{k}) &-H_{\alpha\beta,\downarrow\downarrow}^{\mu\nu *}(-\bm{k})  + E_F
		\end{pmatrix},
		\label{eq:bdgsystem}
	\end{equation}
	with $E_F$ being the Fermi energy while the  associated eigenvalue problem to solve reads:
	\begin{equation}
		\sum_{\beta\mu} H_{\text{BdG}}^{\alpha\beta,\mu\nu}(\bm{k}) \phi_{\beta\mu}(\bm{k}) = E_n(\bm{k}) \phi_{\alpha\nu}(\bm{k}).
		\label{eq:bdg-short}
	\end{equation}
	
	The transformation is canonical and enabled by rewriting 
	\begin{equation}
		c_{\alpha\mu\sigma}(\bm{k}) = \sum_{n}^{'} u_{\alpha\sigma}^n(\bm{k}) \gamma_n + v_{\alpha\sigma}^{n*}(\bm{k}) \gamma_{n}^\dagger, \quad c_{\alpha\mu\sigma}^\dagger(\bm{k}) = \sum_{n}^{'} u_{\alpha\sigma}^{n*}(\bm{k}) \gamma_n^\dagger + v_{\alpha\sigma}^{n}(\bm{k}) \gamma_{n}^\dagger,
		\label{eq:bv-trans}
	\end{equation}
	with
	\begin{equation}
		\{\gamma_n,\gamma_m\} = \{\gamma_n^\dagger,\gamma_m^\dagger\}=0, \quad \{\gamma_n^\dagger,\gamma_m\} = \delta_{nm},
	\end{equation}
	where the prime indicates that the sums run only over the states with positive energy\cite{zhu2016bogoliubov,de1999superconductivity}. 
	This restriction in the sum is done to counteract the doubling of the degrees of freedom originated from the change of basis.

	The eigenvector of Eq.~\eqref{eq:bdg-short} is
	\begin{equation}
		\phi_{i\nu}(\bm{k}) = \begin{pmatrix}
			u_{\alpha\nu\uparrow}(\bm{k})\\ 
			u_{\alpha\nu\downarrow}(\bm{k})\\ 
			v_{\alpha\nu\uparrow}(\bm{k})\\ 
			v_{\alpha\nu\downarrow}(\bm{k})
		\end{pmatrix}.
	\end{equation}
	
	Our numerical procedure to evaluate the various properties of a given material is based on the self-consistency of the charge density, which is also pursued once the electron-phonon coupling is included. In practice, we start the simulations considering the normal metallic phase then we incorporate the Cooper pairing  and repeat the computational runs until convergence. 
	
	\section{Results}\label{sec:results}
	
	In this section we explore the results of simulations done for various Nb structures. Our choice on Nb is motivated by its relatively large superconducting gap, which facilitates the exploration of in-gap states.  
	Simulating low temperatures phenomena requires finer computational meshes, thus for the first investigations we choose high temperatures and large values for the superconducting coupling $\lambda$. We start with bulk Nb, where we recover a superconducting gap opening in the electronic structure around the Fermi energy. Then we explore the case of thin films down to the single monolayer along the [001] or [110] directions. We monitor how the superconducting gap is affected by the self-consistency of the electronic structure calculations. 
	
	\subsection{Bulk superconducting Nb}
	
\begin{figure}[h!]
		\centering
		\includegraphics[width=0.55\linewidth]{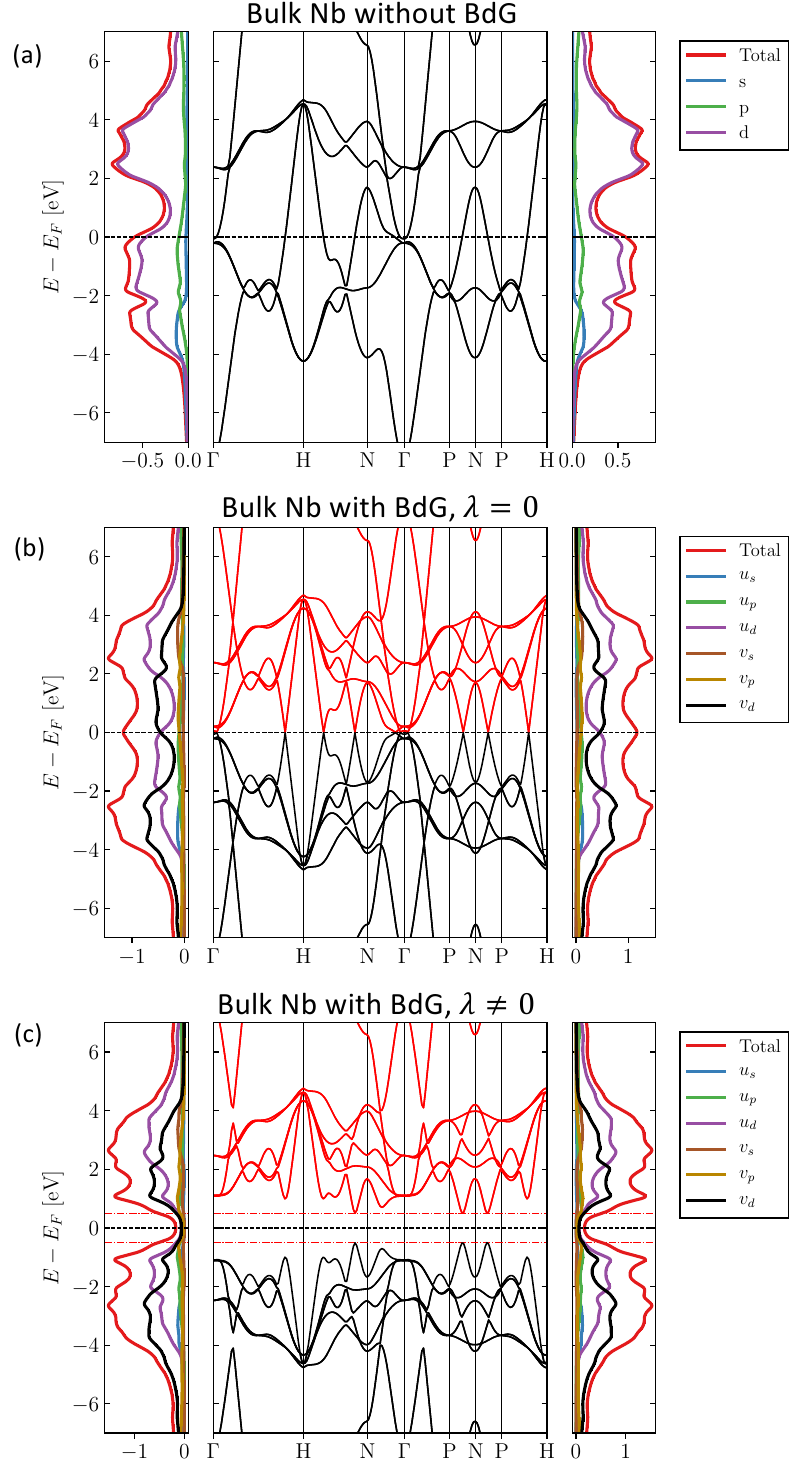}
		\caption{\textbf{Electronic structure of bulk Nb with and without superconductivity.} Three cases are illustrated: \textbf{a} without the BdG formalism only the electron part is shown; \textbf{b} with the BdG formalism but setting the Cooper pair coupling $\lambda$ to zero; \textbf{c} with BdG and $\lambda=\SI{5.44}{\electronvolt}$ in all orbitals. The red colored line evinces the location of the gap as predicted by the self-consistent value $\Delta = \SI{0.497}{\electronvolt}$ and thermal broadening $\eta=\SI{43}{\milli\electronvolt}$. The  band structure is shown in the middle. To the right (left), the DOS of the majority (minority) spin channel is presented. The legend box depicts the colors used to signal the total,  the band resolved DOS and their electron versus hole parts, i.e. the contributions from respectively the $u$ and $v$ components of the eigenvectors.}
		\label{fig:simpleNb}
	\end{figure}

	Nb has a superconducting critical temperature of $T_C = \SI{9.3}{\kelvin}$\cite{Ikushima1969}, which leads to a significant superconducting gap  $2\Delta=\SI{3.8}{\milli\electronvolt}$.
	A rough approximation relates $T_c$ with the size of the gap---at T=0~K---through~\cite{tinkham2004introduction}
	\begin{equation}
		\Delta(T=0) = 1.764\, k_b T_c\,.
		\label{predict}
	\end{equation}\index{Gap equation at $T=0$}
	Large values of $\Delta$ are desirable when investigating states that appear inside the gap. In our simulation procedure, we  start with bulk Nb at high temperatures. In TITAN, the thermal effects enter through the parameter $\eta = k_b T / \pi$, which broadens the Fermi-Dirac function; in this case, we will use $k_b T = \SI{43}{\milli\electronvolt}$ ($\eta = \SI{1}{\milli\rydberg}$). 
	For a reference, this is higher than room temperature, which is at $\SI{26}{\milli\electronvolt}$, but nevertheless, considerably lower than the melting point of Nb $\SI{237}{\milli\electronvolt}$. Nb has a body-centered cubic crystal structure. We make use of the Slater-Koster (SK) parameters from Ref.~\citenum{papabook}, the spin-orbit coupling strength comes from Ref.~\citenum{shanavas}. In Fig.~\ref{fig:simpleNb}(a), we show the density of states of both spin channels (left and right panels), and in the middle the band structure of bulk Nb. The charge occupancy per orbital is $\langle n_s \rangle = 0.70\,,    \langle n_p \rangle = 0.68\,,  \langle n_d \rangle = 3.61.$
	The density of states (DOS) and the band structure agree with the results displayed in Ref.~\citenum{papabook}.
	
	Repeating the calculations by solving the BdG equations but setting $\lambda = 0$ on all orbitals, we naturally recover the metallic band-structure shown in Fig.~\ref{fig:simpleNb}(b), where both the electron and hole parts are illustrated. Both parts are symmetric with respect to the Fermi level. This holds true for the DOS as well. We expect that by allowing for finite coupling $\lambda$, degeneracies occurring at crossing points will be lifted.
	
	In subsection~\ref{sec:tuning}, we describe a rather optimized search method to tune this parameter. From our explorations, we found that: (i) large values of $\lambda$ do not always mean large gaps and (ii) the size of the gap is not a linear function of $\lambda$.  From tabulated values of the electron-phonon $\lambda_{\text{e-p}}$ coupling constant we can evaluate the coupling parameter $\lambda$ utilizing the following relation~\cite{PhysRevB.94.140502}
	\begin{equation}
		\lambda = \lambda_{\text{e-p}}D(E_F)\,,
	\end{equation}
	where $D(E_F)$ is the density of states at the Fermi energy. For example, for bulk Nb with the data from Refs.~\citenum{PhysRevLett.32.1193} we obtain $\lambda \approx \SI{1.3}{\electronvolt}$. In Ref.~\citenum{PhysRevB.94.140502}, the electron-phonon coupling constant is calculated for slabs of Nb(100) and Nb(110) of 3, 6, 9, 12, and 15 layers. For the Nb(110), the results for the system of 9 layers---the slab size closer to the one we will study later---lead to coupling parameters of $\lambda\approx \SI{5.28}{\electronvolt}$ at the surfaces and $\lambda \approx \SI{1.3}{\electronvolt}$ in the middle layer. When plotting the curve associated to $\Delta$ vs. $\lambda$, we obtain a quadratic function, in agreement with Ref.~\citenum{PhysRevB.101.064510}, where the self-consistent Korringa--Kohn--Rostoker method is used. There, the authors found that in order to obtain $\Delta \approx \SI{1}{\milli\electronvolt}$ the coupling parameter must be $\lambda \approx \SI{1.1}{\electronvolt}$ for bulk Nb.
	
	As a first example of our investigations, we set $\lambda_{\mu}=\SI{5.44}{\electronvolt}$.  The occupancies are $\langle n_s \rangle = 0.70317397\,,   \langle n_p \rangle = 0.66860504\,,   \langle n_d \rangle = 3.62822099$, 
	which experience a minor change of the order of $10^{-2}$ with respect to the case where $\lambda$ is set to zero.  The self-consistent value of the superconducting gap parameter is $\Delta = \SI{0.497}{\electronvolt}$, which is obviously gigantic. The resulting band structure and LDOS plots are illustrated in Fig.~\ref{fig:simpleNb}(c).  The red dotted lines delimit the range $E\in [-\Delta,\Delta]$. 
	As a sanity check, it is reassuring that the gap in the band structure matches the self-consistent output value of $\Delta$. The density of states does not show a perfect gap, but decays at the gap region. This effect is due to the thermal broadening induced artificially, which is in this case set to \SI{43}{\milli\electronvolt}.
	
	In subsections~\ref{sec:tuning} and ~\ref{sec:mono110}, we will systematically explore the impact of temperature and coupling strengths, searching for an optimal $\lambda$ that yields the experimental value of $\Delta$.
	
	\subsection{Superconducting gap tuning: single bcc (001) Nb monolayer}\label{sec:tuning}
	
	Explorations to find the appropriate coupling $\lambda$ that results in the desired gap parameter $\Delta$ involve systematic repeated simulations with different parameters.  In this section, we show the typical curves for $\Delta$ as a function of the temperature (via the broadening parameter) and the Cooper pair coupling parameter. To exemplify, we use a 2D system, namely a (001) monolayer of bcc Nb. As in our previous examples, we set the same $\lambda$ for all orbitals, with eight values between \SI{4}{} and \SI{9}{\electronvolt}. We consider forty different temperature points between \SI{300}{\kelvin} and \SI{14000}{\kelvin}. Simulations at lower temperatures are computationally highly expensive since they require an increased number of k-points to be able to capture the narrow peaky structures 
	in the Brillouin zone with lower broadening. All the simulations in this section use the same number of k-points, $\approx2\times 10^6$.
	
	\begin{figure}[h!]
		\centering
		\includegraphics[width=\linewidth]{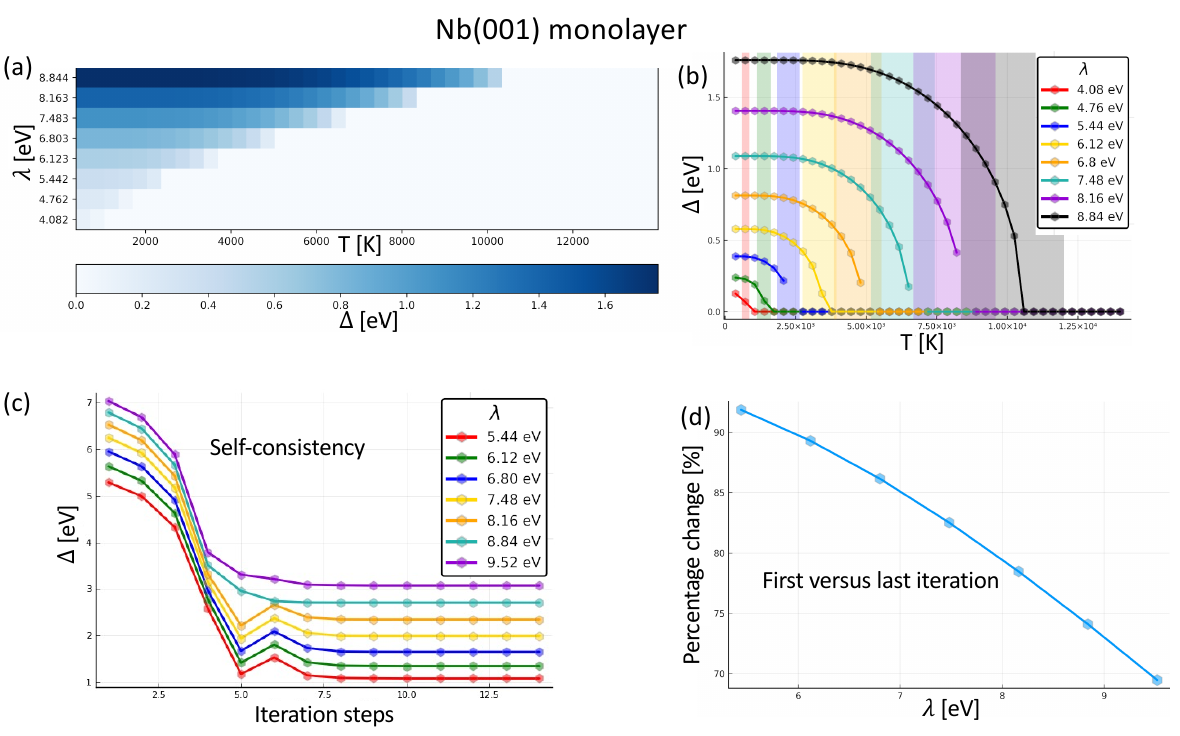}
		\caption{\textbf{Dependence of the superconducting gap of Nb(001) monolayer on various parameters.} \textbf{a} Heat map of the superconducting gap  as function of the coupling $\lambda$ and the thermal  broadening parameter. All curves drop to zero after reaching their corresponding critical temperature. \textbf{b} Another version of \textbf{a} in order to better follow how the superconducting gap evolves against temperature. The bands in color reflect an rough extrapolation to the critical temperature. \textbf{c} Evolution of the superconducting gap  across the self-consistency for several coupling strengths $\lambda$.  \textbf{d} Relative change of the superconducting gap  when comparing the values obtained after one single iteration to those after self-consistency ($(\Delta_{\text{initial}}-\Delta_{\text{final}})/\Delta_{\text{initial}}$) for different values of $\lambda$. 
		}
		\label{fig:imshow}
	\end{figure}
	
	Fig.~\ref{fig:imshow}(a) displays a heatmap with the results of the gap parameter for the different temperatures and couplings. We can see that the gap parameter drops to zero after reaching the corresponding critical temperature for every line. Directing our attention for fixed temperatures (i.e. the ``columns'' in the figure), it is clear that there is also a critical $\lambda_c$ below which $\Delta$ is zero. Throughout the heatmap, $\Delta$ ranges from 0.0 to \SI{1.76}{\electronvolt}.  
	We can analyze the same data from a different perspective, as shown in Fig.~\ref{fig:imshow}(b). There we see that the $\Delta$ vs. $T$ curve has indeed the shape predicted by the BCS theory. For sufficiently low temperatures, $\Delta$ approaches a plateau. Around the critical temperature $T_c$, the self-consistency procedure may become unstable and difficult to achieve a precise value, which lead some curves to miss points in that range. We can proceed to an extrapolation to evaluate the missing points on the basis of the asymptotic behavior of $\Delta$ around the critical temperature~\cite{tinkham2004introduction}
	\begin{equation}
		\Delta(T) \approx 1.74\Delta(0) \sqrt{1-\frac{T}{T_c}}\,.\label{asympgap}
	\end{equation}
	\index{Asymptotic gap equation}
	We used the precedent equation to go from the data in Fig.~\ref{fig:imshow}(b) to the one in Fig.~\ref{fig:imshow}(a). 
	The BCS theory predicts that, at zero Kelvin, the relation between $\Delta$ and $T_c$ is the one in Eq.~\eqref{predict}. 
	Experimentally the reported values range from $2\Delta = 3.2 k_b T_c$ to $2\Delta = 4.6 k_b T_c$, we signaled this range with the colored bands in Fig.~\ref{fig:imshow}(b). This plot establishes the aforementioned statement regarding the non-linearity of $\Delta$ with respect to $\lambda$. 
	
	For a wide range of simulations, it is desirable and recommendable to place the superconducting system well below $T_c$, at a point in the plateau of the $\Delta$ vs. $T$ curve. The superconducting state is frailer at the vicinity of the critical temperature, and internal changes in the material can perturb the properties of interest. 
	
	\subsection{Renormalization after self-consistency: single bcc (001) Nb monolayer}
	
	A common problem in simulating superconductors interfaced with normal conductors   
	is the artificially increased size of the gap in order to facilitate the exploration of emerging in-gap states and it is rather common to proceed to single-iteration simulations, which assumes the electronic structure is not significantly altered in comparison to the case, where instead one proceeds to self-consistency. The importance of self-consistency was highlighted in Ref.~\citenum{Li2016}. In this subsection, we address this aspect of simulation of superconducting materials by comparing the size of the gap obtained after one iteration (starting from the converged electronic structure of the metallic phase) with respect to the one obtained after full self-consistency. In Fig.~\ref{fig:imshow}(c), we single out the value of the superconducting gap parameter through several simulations with different coupling parameters at a temperature of $\SI{43}{\milli\electronvolt}$. Our algorithm, based on the Powell method~\cite{Powell1964} can halt the simulations when no improvement is detected in the last five steps. All shown cases converged without a problem.
	
	In Fig.~\ref{fig:imshow}(d) we portray the percentage change of the gap parameter  ($(\Delta_{\text{initial}}-\Delta_{\text{final}})/\Delta_{\text{initial}}$) between the first- and last-iteration calculated $\Delta$ for different $\lambda$. We note that the change is bigger for smaller values of $\lambda$. 
	
	So far, we have been discussing the numerics of the simulations assuming both large superconducting gaps and large electronic temperatures.  
	In section~\ref{sec:mono110}, we drop $T$ to realistic values associated to a superconducting Nb layer. 
	Additionally, we explore the impact on the orbital-dependent gaps $\Delta_\mu$.
	
	\subsection{Superconducting Nb (110) monolayer}\label{sec:mono110}
	
	In this subsection we consider the cryogenic regime, which calls for computationally heavier simulations. As we hinted in previous subsections, sinking the temperature forces us to increase the number of k-points for each simulation to be able to capture the narrow states. Since these calculations will be more intensive, we will leave aside the testing structures we were using previously---bulk Nb and Nb (100) monolayers---and move to our main interest, which is Nb (110) films, heavily explored in various recent experiments. 
	
\begin{figure}[!h]
		\centering
		\includegraphics[width=1.0\linewidth]{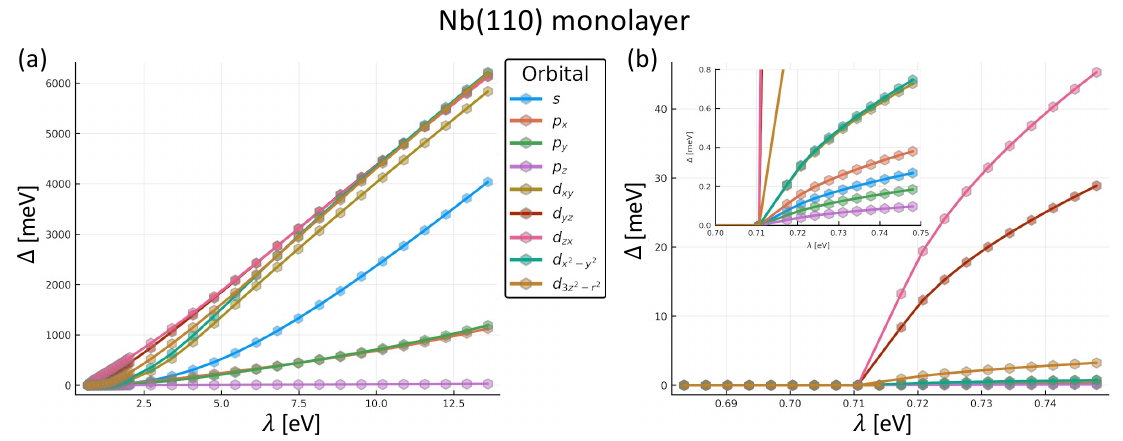}
		\caption{\textbf{Orbital-resolved superconducting gap for Nb(110) monolayer.} \textbf{a} $\Delta_\mu$ grows considerably for all the orbital except $p_z$. \textbf{b} Two close ups to the region where the material transitions from the non-superconducting to the superconducting phase. Simulations performed at $\SI{4.56}{\kelvin}$.}
		\label{fig:monoDeltas}
	\end{figure}

	We assume a thermal broadening of $\SI{4.27}{\milli\electronvolt}$ (equivalent to a temperature of approximately $\SI{4.56}{\kelvin}$), shallow enough to be in the plateau of the $\Delta$ vs. $T$ curve for all the $\lambda$ we used. We consider a non-uniform grid with 57 points for $\lambda$, ranging from $\SI{0.68}{\electronvolt}$ to $\SI{13.6}{\electronvolt}$, with a higher resolution for lower energies. In Fig.~\ref{fig:monoDeltas}(a), we display the results of the orbital-dependent $\Delta_\mu$ for the entire range. Obviously, some orbitals are more responsive to the coupling $\lambda$. This is affected by the strength of the latter coupling. 
	We notice an increment in $\Delta$ as we enlarge the coupling strength. $\lambda > \SI{2.5}{\electronvolt}$ induces gaps of the order of $\SI{1e2}{\electronvolt}$ or larger, high enough to be considered unrealistic.
	
	In Fig.~\ref{fig:monoDeltas}(b), we focus only on the low energy range $\lambda\in[\SI{0.5}{\electronvolt},\SI{0.75}{\electronvolt}]$, where the growth of $\Delta(\lambda)$ is more controlled. In the inset, we also restrict the values on the $y$-axis from $0$ to \SI{0.8}{\electronvolt} to resolve better the smaller gap parameter for some of the orbitals. We observe that the $p_z$ component is still quite small; this is a consequence of the shape of this orbital and the geometry of the monolayer. The orientation of the $p_z$ orbital is out of the plane, and this means its in-plane component is zero, and it is hard for superconducting electrons in this orbital to interact with others.

	The gap is opened around the Fermi surface when the $u$ and $v$ states mix due to the off-diagonal components of the Hamiltonian; this means that the gap must come from the bands that cross $E_F$. Hence, the gap of the system comes from the minimum value of $\Delta_\mu$ among orbitals whose bands traverse the Fermi level. Once we profile these bands, we can execute any searching algorithm---e.g., binary search, or nested intervals methods---until we find an appropriate $\lambda$. As a rule of thumb, each significant figure in $\Delta$ requires two on $\lambda$; In other words, having $\Delta$ up to three significant digits typically requires knowing $\lambda$ to six. Refining can be very expensive. In our case, $\lambda = \SI{1.37}{\electronvolt}$ induces $\Delta=\SI{1.4}{\milli\electronvolt}$.
	
	In the next subsection, we explore the case of thin films of Nb (110) at $\SI{4.56}{\kelvin}$ and analyse the evolution of the superconducting gap parameter across the material.
	
	\subsection{Superconducting slab}
	
	It is difficult to predict if the coupling strength $\lambda$ that works well for one given system still opens a reasonable gap for a different one. We started with the value found in the previous section, $\lambda = \SI{1.37}{\electronvolt}$, which opened a gap $\Delta=\SI{1.4}{\milli\electronvolt}$ for a single monolayer. It turns out, however, that as soon as the Nb film thickness is larger than 5 layers, a large coupling is required to open a gap. In the remained study, we assume $\lambda=\SI{2.72}{\electronvolt}$ at a temperature of $\SI{4.56}{\kelvin}$. 
	
\begin{figure}[h!]
		\centering
		\includegraphics[width=1.0\linewidth]{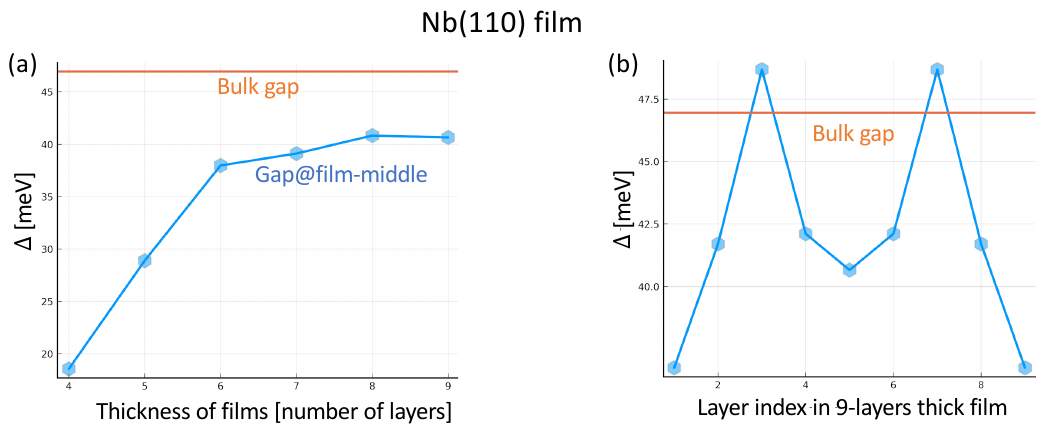}
		\caption{\textbf{Superconducting gap for Nb(110) films.} \textbf{a} Superconducting gap at the centers of Nb(110) films of  different thicknesses.  \textbf{b} Layer-resolved superconducting gap for a nine-layer-thick Nb(110) film. The orange line represents the gap (\SI{46.9}{\milli\electronvolt}) associated to bulk Nb assuming $\lambda=\SI{2.72}{\electronvolt}$ and a temperature of $\SI{4.56}{\kelvin}$.}
		\label{fig:gapsSlab}
	\end{figure}

	In Fig.~\ref{fig:gapsSlab}(a), we show the magnitude of the gap in the central layer of the films as a function of the slab thickness. The orange line corresponds to the gap parameter of the bulk Nb for $\lambda=\SI{2.72}{\electronvolt}$ at the same temperature. One notices that the thicker the film gets, the closer the gap is to the one obtained for the bulk, a behavior comparable to what has been reported in Ref.~\citenum{Kodama1983}. 
	
	The superconducting gap is not homogeneous across the Nb films. An example is shown in Fig.~\ref{fig:gapsSlab}(b) for a film with a thickness of 9 layers. The gap in the middle of the film is larger than at the surface and it seems to oscillate as a function of position, as we would expect for wave-like particles in a confined system. A lower gap at the surface is expected since the electrons there have fewer opportunities to couple in Cooper-pairs due to their lack of neighbors from one side. We must also consider surface effects; for example, we have around 1.5 electrons less than the inner parts. Electrons move from the surfaces to the middle layers. Interestingly some of the layers in the film can have a larger gap than the bulk one. This indicates that the dimensionality of a material, can be utilized to control the superconducting gap due to confinement effects.
	
	\iffalse 
	Among the layers in the slab, the central one should have properties closer to the bulk material since its distance to the surfaces is larger and it is less prone to be affected by surface effects. In Fig.~\ref{fig:bigdosslab}, we display the local density of states of atoms in the middle layer for the $u$ and $v$ components of the eigenvectors. As we observe, the d-bands are predominant; naturally, as they fit 10 electrons.
	\begin{figure}[h!]
		\centering
		\includegraphics[width=0.9\linewidth]{Pictures/largerange.png}
		\caption{Local density of states at the central layer of a nine-layer-thick Nb(110) for an energy range for -8 to 8~eV. The simulation uses a coupling strength of $\lambda=2.72$~eV and a temperature of $4.56$~K.
		}
		\label{fig:bigdosslab}
	\end{figure}
	In Fig.~\ref{fig:bigdosslabzoom}, a close up of the local density of states in the vicinity of the Fermi level highlights the gap around zero. The self-consistent value of $\Delta=40.6$~meV matches the region where the number of states drops to zero.
	
	\begin{figure}[!htb]
		\centering
		\includegraphics[width=0.9\linewidth]{Pictures/slabldoszoom.pdf}
		\caption{Local density of states at the central layer of nine-layer-thick Nb(110) for an energy range for -0.5 to 0.5~eV. The red vertical lines mark the limits of the gap at $\Delta=40.6$~meV. The simulation uses a coupling strength of $\lambda=2.72$~eV and a temperature of $4.56$~K.
			% \samir{Always add Lambda and temperature}\uriel{added}
			%\uriel{added}\samir{ok}
		}
		\label{fig:bigdosslabzoom}
	\end{figure}
	\fi
	
	\section{Conclusions}\label{sec:conclusions}
	
	After briefly introducing the tight-binding method that we developed to tackle superconductivity within the Bogoliubov-de Gennes formalism, we presented the results of our simulations on a material that is currently heavily utilized to explore the physics of superconductivity. We addressed the case of bulk and thin films of Nb along different directions. A major issue in this field from the computational point of view is to choose the right Cooper pair coupling that opens the desired superconducting gap since the calculations are extremely expensive to resolve gaps of a few \SI{}{\milli\electronvolt} or even smaller. We analysed how the superconducting order parameter correlates with both the Cooper pair coupling and temperature, and explored the impact of self-consistency on the emerging orbital-dependent superconducting gap. In the thin film geometry, we unveiled the possibility of engineering the magnitude of the gap via confinement effects. Indeed, the superconducting gap is found to be layer-dependent, either smaller or larger than the values found in the bulk phase. 
	
	\begin{acknowledgments}
		The authors acknowledge funding provided by the Priority Programmes SPP 2137 ``Skyrmionics'' (Projects LO 1659/8-1) of the Deutsche Forschungsgemeinschaft (DFG). 
		We acknowledge the computing time provided through JARA on the supercomputers JURECA\cite{Krause_2018}. Simulations were also performed with computing resources granted by RWTH Aachen University under project jara0189 and p0020362.
	\end{acknowledgments}
	
	\bibliography{bibliography}% Produces the bibliography via BibTeX.
	
\end{document}